\documentclass[conference]{IEEEtran}
\IEEEoverridecommandlockouts

\usepackage{cite}
\usepackage{amsmath,amssymb,amsfonts}
\usepackage{algorithmic}
\usepackage{graphicx}
\usepackage{textcomp}
\usepackage{xcolor}
\usepackage{subcaption}
\usepackage{csquotes}
\usepackage{url}

\usepackage{hyperref}
\usepackage[hyphenbreaks]{breakurl}
\def\BibTeX{{\rm B\kern-.05em{\sc i\kern-.025em b}\kern-.08em
    T\kern-.1667em\lower.7ex\hbox{E}\kern-.125emX}}
\begin{document}
\title{Partial Symbol Recovery for Interference Resilience in Low-Power Wide Area Networks\\
\thanks{
\IEEEauthorrefmark{1}Both authors contributed equally to this work.
\IEEEauthorrefmark{3}Shuai Wang is the corresponding author.}
}
\author{
\IEEEauthorblockN{Kai Sun\IEEEauthorrefmark{2}\IEEEauthorrefmark{1}~~~~~
Zhimeng Yin\IEEEauthorrefmark{4}\IEEEauthorrefmark{1}~~~~~
Weiwei Chen\IEEEauthorrefmark{2}~~~~~
Shuai Wang\IEEEauthorrefmark{2}\IEEEauthorrefmark{3}~~~~~
Zeyu Zhang\IEEEauthorrefmark{2}~~~~~
Tian He\IEEEauthorrefmark{2}} \\
\IEEEauthorblockA{\small \IEEEauthorrefmark{2}
%School of computer science and engineering, 
Southeast University, %China,
\IEEEauthorrefmark{4}City University of Hong Kong\\
\{sunk, chen\_weiwei, shuaiwang, 213161182,tianhe\}@seu.edu.cn,
zhimeyin@cityu.edu.hk
}
}
\maketitle

% \maketitle
\begin{abstract}

Recent years have witnessed the proliferation of Low-power Wide Area Networks (LPWANs) in the unlicensed band for
various Internet-of-Things (IoT) applications.
Due to the ultra-low transmission power and long transmission duration, LPWAN devices inevitably suffer from high power Cross Technology Interference (CTI), such as interference from Wi-Fi, coexisting in the same spectrum. To alleviate this issue, this paper introduces the Partial Symbol Recovery (PSR) scheme for improving the CTI resilience of LPWAN. We verify our idea on LoRa, a widely adopted LPWAN technique, as a proof of concept.

At the PHY layer, although CTI has much higher power, its duration is relatively shorter compared with LoRa symbols, leaving part of a LoRa symbol uncorrupted. Moreover, due to its high redundancy, LoRa chips within a symbol are highly correlated. This opens the possibility of detecting a LoRa symbol with only part of the chips. By examining the unique frequency patterns in LoRa symbols with time-frequency analysis, our design effectively detects the clean LoRa chips that are free of CTI. 
This enables PSR to only rely on clean LoRa chips for successfully recovering from communication failures.
We evaluate our PSR design with real-world testbeds, including SX1280 LoRa chips and USRP B210, under Wi-Fi interference in various scenarios.
Extensive experiments demonstrate that our design offers reliable packet recovery performance, successfully boosting the LoRa packet reception ratio from 45.2\% to 82.2\% with a performance gain of 1.8$\times$.

\end{abstract}
\begin{IEEEkeywords}
LPWAN, LoRa, Wireless interference
\end{IEEEkeywords}

\section{Introduction}

The proliferation of IoT applications, e.g., smart city and environmental monitoring \cite{mekki2019comparative}\cite{neumann2016indoor}, call for Low-Power Wide Area Network (LPWAN). 
It enables IoT devices to reach a communication distance of a few kilometers, in contrast to the widely adopted wireless technologies (e.g., Wi-Fi, ZigBee, and Bluetooth) with limited communication distances. 
With LPWANs, low-power and low-cost IoT devices could be deployed ubiquitously at any place in a city, 
where they report sensing data to the central servers at a very low frequency for long-time operation.

To enable low-power and long-range communication, several communication techniques have been introduced, including
LoRaWAN \cite{LoRawan_ref}, SigFox \cite{sigfox_ref}, NB-IoT \cite{ratasuk2016nb} and LTE-M \cite{lte_m_ref} .
Among them, NB-IoT and LTE-M require licensed spectrum,
while LoRaWAN and SigFox propose to utilize the unlicensed Industrial, Scientific and Medical (ISM) band. 
Specifically, techniques operating in the licensed spectrum have dedicated spectrum resources but they require permissions at extra cost. In contrast, LPWANs in the unlicensed spectrum remove this spectrum cost, enabling 
the wide deployment across various countries. 
However, this also leads to spectrum competition with wireless technologies~\cite{dongare2018charm} and inevitable performance degradation of IoT applications\cite{orfanidis2017investigating}.

Given the long-range communication nature of LPWANs, where the communication range could reach 10 km, the signal power at the receiver side is extremely low, e.g., -132 dBm in LoRa \cite{LoRa_power_ref}.
This ultra-low-power communication could be easily destroyed by ongoing cross-technology interference. 
For example, in the 915 MHz ISM band, the LoRa signal could be corrupted by technologies like 802.11 ah.
As for the 2.4 GHz LoRa, introduced by Semtech in 2017, it suffers from severe wireless interference, including ZigBee, Bluetooth, and high-power Wi-Fi \cite{LoRa_wifi_ref}\cite{LoRa_bluetooth_ref}. 
In addition to the power asymmetry, the long transmission duration of LoRa further aggregates the chances of collisions. For example, a LoRa packet could last as long as 8,000 ms, while a normal inter-packet duration in Wi-Fi is less than 3ms\cite{zhao2018spatial}.

To alleviate this severe CTI problem, given the explosive growth of IoT devices of different types, this paper introduces a partial symbol recovery design for protecting low-power LPWAN techniques in the unlicensed spectrum. Without loss of generality, we present our design on the 2.4 GHz LoRa, while its design principle could also be applied to other scenarios. 

To enhance the CTI protection, our design first examines the unique features of LoRa communication under CTI. 
Due to the low transmission rates, a LoRa symbol is relatively much longer than the duration of CTI, leaving partial signals correct. 
By utilizing this unique phenomenon at the PHY layer, our design examines the frequency-time pattern for finding correct LoRa chips within corrupted LoRa symbols. 
Although this idea seems to be straightforward, it is a non-trivial task to achieve this goal for LPWAN.
First, CTI is dynamic and changing all the time. There could be multiple interference transmissions with different power/duration from different wireless technologies within one LoRa symbol.
Second, designed for long-range communication, LoRa communication has extremely low power, even smaller than background noises.
To address these issues for accurate CTI identification, our Partial Symbol Recovery (PSR) proposes a two-step partial symbol recovery technique. In step 1, coarse-grained symbol location, PSR examines the special frequency-time patterns in LoRa by computing the Short-time Fourier transform (STFT). This is motivated by the difference that LoRa communication has concentrated spectral components, while CTI does not have this feature. After STFT, PSR further utilizes max pooling and computes the ratio between dominating frequency component and average.
In step 2, fine-grained symbol detection, PSR examines the computed ratio for accurately identifying the clean LoRa chips free of CTI. After this, it recovers the corrupted LoRa symbol through correlation detection.

Specifically, our design has the following contributions:
\begin{itemize}

\item To the best of our knowledge, this is the first partial symbol recovery technique in LPWAN under wireless interference. Our design identifies the clean signal within a corrupted LPWAN symbol for accurate symbol recovery. It does not require prior knowledge of the interference, while also avoiding demodulation of the interference in contrast to existing related work\cite{balanuta2020cloud}.

\item To realize the partial symbol recovery, we propose a two-step technique. First, our design computes the short-time Fourier transform to analyze the frequency-time pattern. Then it computes the max pooling and frequency component to the average ratio for roughly locating the regions of the transmitted LoRa symbol. Second, it identifies the clean LoRa chips that are not corrupted by interference and further recovers the corrupted LoRa symbol based on clean LoRa chips.

\item We implement the prototype with LoRa SX1280 chip and USRP B210, and evaluate our design under Wi-Fi, the dominating interference source in 2.4GHz. Extensive results demonstrate that our PSR design improves the packet reception ratio from 45.2\% to 82.2\%, achieving a performance gain of 1.8$\times$.

\end{itemize}

\section{Motivation}
This section presents the motivation of this paper, i.e., the deployment of LoRa in the unlicensed spectrum, and the related impacts of CTI.

\subsection{LoRa in the Unlicensed Spectrum}
To offer long-range and low-power communication, Semtech introduces LoRa with the Chirp Spread Spectrum (CSS) in the sub 1GHz. 
By spreading a LoRa symbol to a long sequence of signals with the dedicated frequency pattern, LoRa communication has a high processing gain, so that it is robust to environmental noises for achieving a long communication distance, e.g., 10 km in the rural areas. 
Due to the different regulations on the ISM band, the operation frequencies of LoRa are different. For example, the ISM band is 915 MHz for North America, while this is 868 MHz in Europe. These diverse regulations lead to extra hardware complexity for the low-cost LoRa devices.

To resolve this practical issue, Semtech recently introduces the 2.4GHz LoRa with commodity chips SX 1280 and SX 1281 \cite{LoRawan_ref}, since the 2.4GHz ISM band is commonly available all over the world. By shifting the operational frequency to 2.4 GHz, enables a LoRa chip to work seamlessly across various regions like North America, Europe and China, and the corresponding fast deployment of LoRa. Specifically, the 2.4 GHz LoRa has very similar modulation and demodulation techniques with the sub-1GHz LoRa. For example, the 2.4GHz still relies on CSS to enable a communication range of more than a few kilometers. In contrast to the existing popular technologies (Wi-Fi, Bluetooth and ZigBee) with a range of usually less than 100 m, LoRa significantly boosts the communication range for enabling potential IoT applications.

\begin{figure}[!htbp]
\centering
\includegraphics[width=0.4\textwidth]{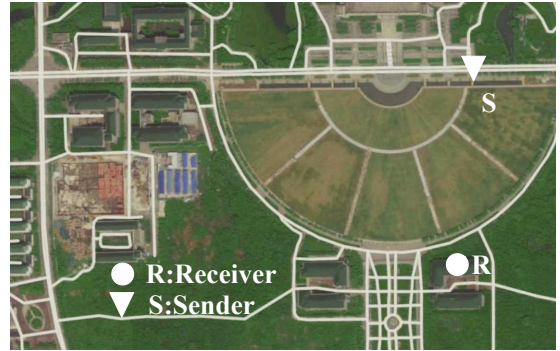}
\caption{Experiment scene}
\label{fig:outdoor_scenario}
\vspace{-6mm}
\end{figure}

\begin{figure}[!htb]
\centering
  % include first image
  \includegraphics[width=0.4\textwidth]{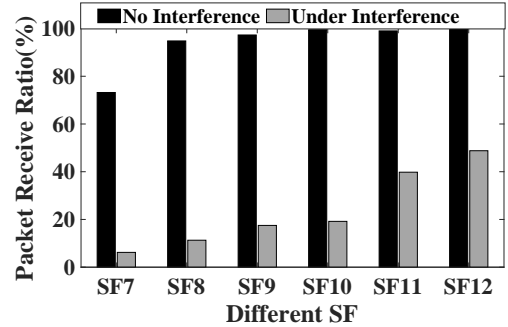}
  \caption{Packet reception ratio comparison}
  \label{fig:prr_dif_sf_tra}
  \vspace{-6mm}
\end{figure}

\subsection{LoRa under CTI}

Since LoRa operates in the unlicensed spectrum, it inevitably needs to compete with other wireless protocols for the limited spectrum resources. This spectrum competition leads to corrupted LoRa transmissions. For example, in the 915 MHz ISM band, LoRa could be affected by the 802.11 ah. As for the well-known 2.4 GHz, which is crowded with various popular wireless protocols, e.g., Wi-Fi, ZigBee and Bluetooth, LoRa suffers from strong performance degradation due to CTI. In addition, LoRa generally has a long transmission duration, while a single LoRa packet could last as long as 270-8000ms. This extremely long on-air time further aggravates the chances of LoRa corruption, given the existing crowded 2.4GHz ISM band. 

In order to prove the influence of the interference on the low SNR LoRa signal, outdoor experiments are conducted on playgrounds and on campus roads.
We utilize the SX1280 LoRa node to transmit 1000 packets in different Spreading Factor(SF) from S where is 200+m away from R shown in Figure \ref{fig:outdoor_scenario}, and the SNR is close to -10dB. We conducted two sets of experiments in the middle of the night and in the morning. In the middle of the night, no Wi-Fi device was connected to the Wi-Fi AP and there was negligible interference. In the morning, more than five wireless devices were connected to the same Wi-Fi AP, leading to high Wi-Fi interference. 
Figure \ref{fig:prr_dif_sf_tra} shows that the packet reception ratio of standard LoRa, which adopts Hamming code with a code rate 4/8. It is obvious that
 standard LoRa performs wells when there is no interference - the receiver
 successfully decoded 73.2\% packets in SF7 and almost 100\% packets in larger SF.
On the contrary, under CTI the packet reception ratio of standard LoRa is only
 6.5\% packets in SF7. As SF increases, the packet receive ratio increases slowly, and all packet receive ratios under different SF are lower than 50\% under CTI. This series of experiments demonstrate the serious impacts of CTI on the LoRa communication reliability.

 \begin{figure}[ht!]
 \centering
 \includegraphics[width=0.48\textwidth]{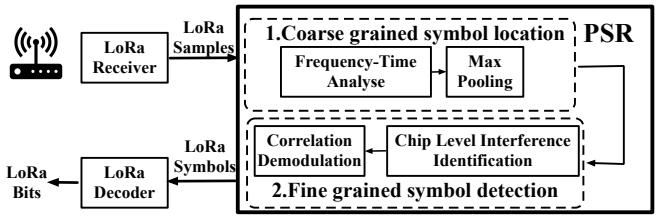}
 \caption{Overview of PSR}
 \label{fig:overview_psr}
\vspace{-4mm}
 \end{figure}

\section{System Overview}

\subsection{Overview}

PSR(Partial Symbol Recover) is a PHY layer software solution that recovers valid symbols that would normally be lost due to interference. It is applicable to losses packets from strong interference in a low SNR environment that are often the case in highly dense areas. Recover the interfered symbol could allow LoRa clients to transmit at higher data rates and at lower powers to preserve battery.

Our design is implemented at the LoRa receiver, and is compatible with all existing LoRa transmitters. By elaborately utilizing the features of the LoRa PHY layer, our design effectively achieves reliable protection performance, even in scenarios where interference is very high. The overall design architecture of PSR is shown in Figure \ref{fig:overview_psr}. Each LoRa symbol will go through a two-stage detection procedure.  

1. Coarse-grained symbol location: this stage is used for analyzing the time-frequency behavior of a LoRa symbol. After performing STFT, Max Pooling and Frequency component to average ratio operation are performed to suppress interference influence. Finally, Frequency component normalization is used to trace the frequency changes over time. 
Intuitively, when there is no CTI the frequency components stay quite stable over time; otherwise, the frequency components will change drastically at the time when CTI occurs. 

2. Fine-grained symbol detection firstly use the normalized time-frequency components to identify whether a chip is interference free or not. It then collects all clean chips and uses correlation detection to decode the LoRa symbol.

\subsection{Challenges}
To reliably protect the low-power and low-cost LoRa against various high-power CTI, how to select clean chips from the polluted ones is crucial. Several challenges regard to this issue are listed as follows.

\begin{itemize}
\item \textbf{An unified approach for handling CTI}. A LoRa symbol can be interfered by one or multiple packets with different power and transmission duration. In addition, these wireless packets could
 belong to different wireless technologies
(e.g. Wi-Fi, ZigBee, BLE or combinations of them), while signals from these technologies can not be decoded at the LoRa receiver. The heterogeneity makes it difficult to identify whether a LoRa chip has been interfered by CTI or not.

\item \textbf{Low SNR signal detection}. Besides CTI, a LoRa packet suffers from strong noise due to its low transmission power and long transmission range. Low SNR implies that noise will pose another challenge when locating CTI within a LoRa symbol. 
\end{itemize}

\par

\begin{table}[thb!]
\newcommand{\tabincell}[2]{\begin{tabular}{@{}#1@{}}#2\end{tabular}}  %导言区
  \centering
  \begin{tabular}{|c|c|}
\hline
Symbol & \tabincell{c}{Notation} \\
\hline
$x_s(n)$ & \tabincell{c}{The $n$th chip of the $s$th LoRa symbol \\ at the receiver side(before downchirp)} \\
\hline
$l_s$ & \tabincell{c}{The number of chips in the $s$th symbol}\\
\hline
$l_{STFT}$ & \tabincell{c}{The number of Non-zero chip in a STFT window}\\
\hline	
$l_{pool}$ & \tabincell{c}{The number of chip in a Max Pooling window}\\
\hline	
% $s$ & \tabincell{c}{The $s$th symbol}\\
% \hline
% $n$ & \tabincell{c}{At the $n$th chip}\\
% \hline
$r_s$ & \tabincell{c}{The $s$th demodulated LoRa symbol}\\
\hline
$\tilde{x}_s$ & \tabincell{c}{The $n$th chip of the $s$th LoRa symbol at the receiver \\ side (after downchirp),$\tilde{x}_s =x_s(n)e^{-j2\pi (\frac{n^2}{2}-\frac{n}{2})}$}\\
\hline
$y_s(m)$ & \tabincell{c}{The $m$th frequency component after FFT}\\
\hline	
$z_s(\tau ,m)$ & \tabincell{c}{The $m$th frequency component when STFT is performed \\ for the $s$th symbol with Hann window centered around \\ the $\tau$th chip}\\
\hline
\end{tabular}

\caption{Notations}
\vspace{-4mm}
\end{table}

\section{Design}

%\subsection{Symbol Summary}

\subsection{Background}
This section introduces the background of LoRa modulation, demodulation, and its reliability against interferences and background noises. For easy reading, we summarize the key notations used in this paper, as shown in Table I.

\begin{figure}[!htb]
\centering
\includegraphics[width=0.4\textwidth]{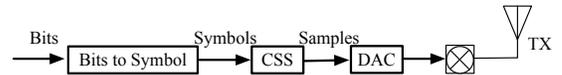}
\caption{LoRa modulation procedures}
\label{fig:background_modulation}
\vspace{-2mm}
\end{figure}

\noindent\textbf{LoRa modulation}
To offer long-range communication, Semtech introduces Chirp-Spread-Spectrum (CSS) technique
while the modulation procedures follow Figure \ref{fig:background_modulation}.
The CSS spreads the input LoRa symbol to a long sequence of related signal samples, for introducing resilience against wireless fading and noises. To allow the trade-off between the transmission speed and the transmission reliability, the SF is defined. Specifically,
 the LoRa transmitter first takes a LoRa symbol with SF bits, and then spreads this LoRa symbol to $2^{SF}$ chips with the following equation
\begin{equation}\label{LoRachip}
x_s(n)=e^{-j2\pi(f_0+ks'+kn)\frac{n}{N}}
\end{equation} 
where $f_0$ is the starting frequency,
$n$ is the chip number within this LoRa symbol, $s'$ is the transmitted symbol, and $k=\frac{2^{SF}}{BW}$ ($BW$ is the LoRa bandwidth).

\begin{figure}[!htb]
\centering
\includegraphics[width=0.4\textwidth]{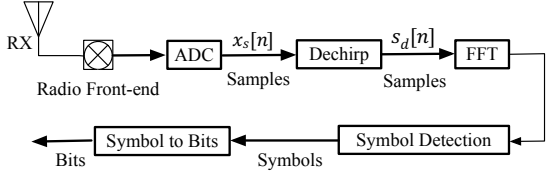}
\caption{LoRa demodulation procedures}
\label{fig:background_demodulation}
\vspace{-2mm}
\end{figure}

\noindent\textbf{LoRa demodulation}
Figure \ref{fig:background_demodulation} depicts the demodulation procedures for a LoRa receiver.
With the received signal $x_s(n)$, the LoRa receiver processes it in the Dechirp procedure.
Specifically, the LoRa receiver first generates a DownChirp, and then multiplies $x_s(n)$ with DownChirp in the Dechirp procedure, $\tilde{x}_s =x_s(n)e^{-j2\pi (\frac{n^2}{2}-\frac{n}{2})}$.
\begin{equation}\label{LoRachip}
r_s=argmax_{0\leq m \leq l_s-1} y_s(m)
\end{equation} 
Eq.\ref{LoRachip} shows the operation of demodulation. After computing the FFT of this sequence, the LoRa demodulator converts the FFT result to each symbol, and picks the symbol with the maximum $|FFT|$ as the demodulation result.

\noindent\textbf{LoRa reliability}
With CSS, LoRa spreads an SF-bit input symbol to $2^{SF}$ chips, thus achieving a processing gain of $\frac{2^{SF}}{SF}$ and adding significant resilience for reaching long communication range. 

Since the background noises are generally weak, CSS effectively adds an additional link budget to the weak LoRa signal, for reaching long distances, e.g., kilometers away.

Although effective against background noises, it struggles to combat CTI, which usually has significantly larger power.
For example, designed for offering high-speed data transfer service, Wi-Fi is known to have very large power, which could be dozens of dB larger than LoRa, thus leading to corrupted LoRa communication as examined in Section II.

In addition, due to the low transmission rate, the transmission duration of a LoRa symbol is generally very long. For a single LoRa symbol, the symbol duration could be as long as 0.625-20ms, depending on the value of SF, while the packet on-air time could reach 20-2000 ms. This is significantly longer than the packet duration of the popular wireless technologies in the ISM band, such as Wi-Fi, ZigBee and Bluetooth, which is generally less than 1 ms.
Given the crowded spectrum in the 2.4GHz, it is common that a LoRa packet with a long transmission duration collides with other ongoing wireless communication.

However, the long duration of LoRa transmission also introduces the possibility of fighting against the CTI. Since a LoRa symbol is generally longer than the duration of CTI, it has partially correct signals free from CTI. Our design benefits from this existing opportunity for recovering corrupted LoRa symbols, even when they have significantly smaller energy, e.g., -30 dB, thus enhancing the performance of LPWAN.

\subsection{Coarse-grain symbol location}

Before recovery, corrupt symbols need to be processed in order to detect which sections of the symbol are corrupt. The goal is to output a vector of clean chips the same size as the number of chips in the symbol. More clean chips are more likely to accurately restore the correct symbol. To extract as much reliable information as possible, each time we use STFT windows of different sizes to extract clean chips to the vector. 

\begin{figure*}[ht]
\begin{subfigure}{.33\textwidth}
  \centering
  % include second image
  \includegraphics[width=\linewidth]{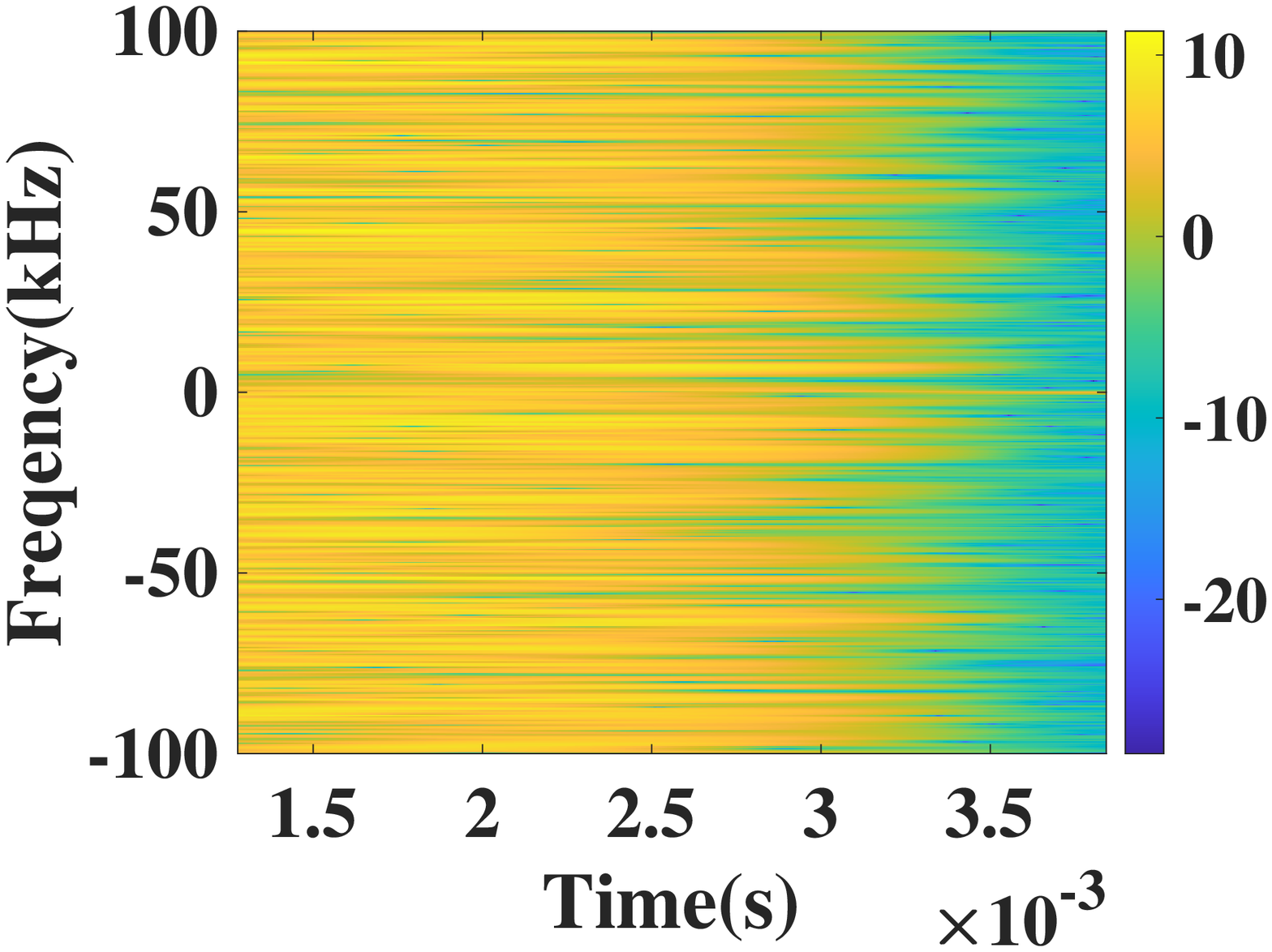}
  \caption{$l_{STFT}=\frac{l_s}{2}$}
  \label{fig:win_size_1_2}
\end{subfigure}
\begin{subfigure}{.33\textwidth}
  \centering
  % include second image
  \includegraphics[width=\linewidth]{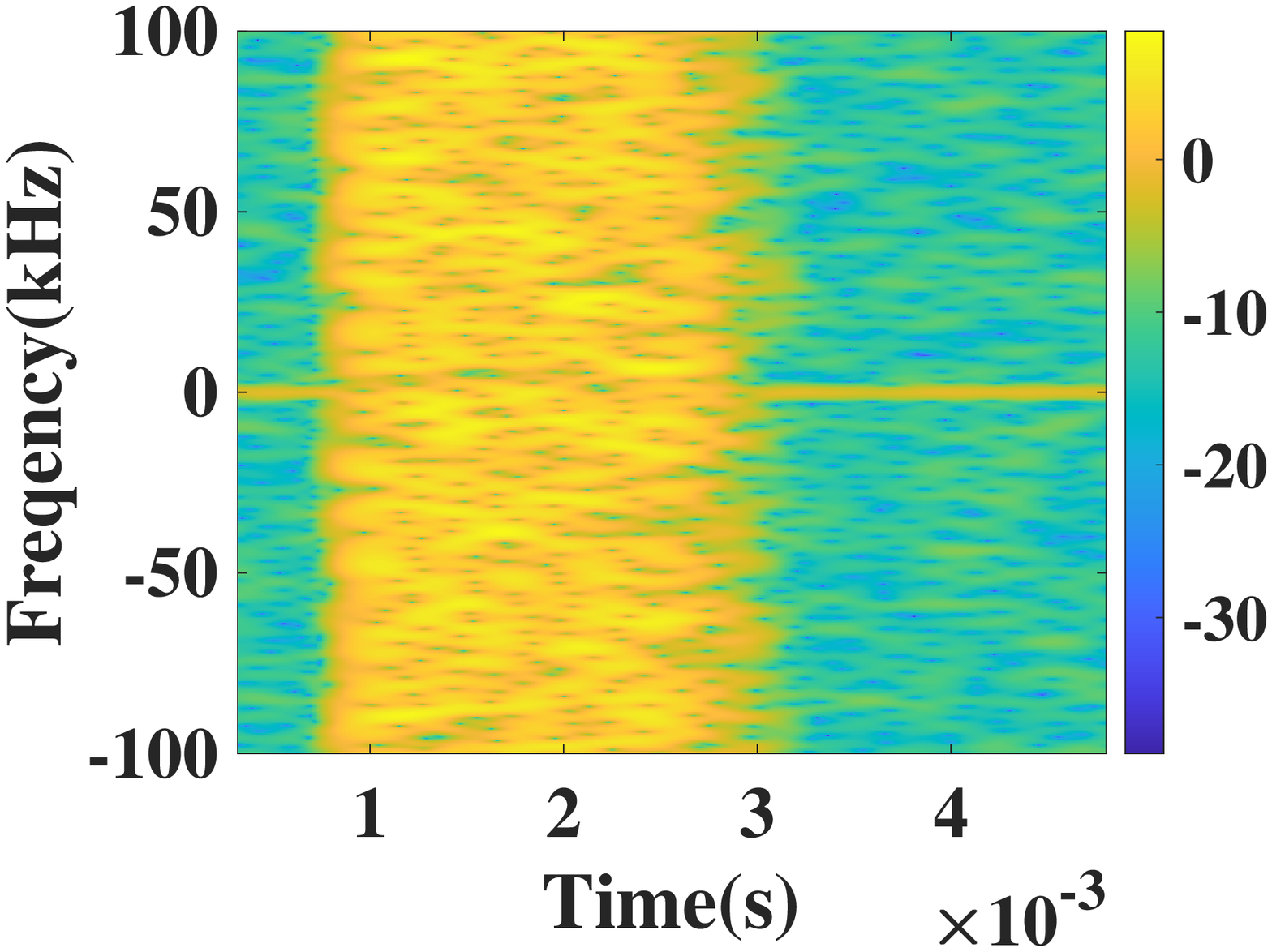}
  \caption{$l_{STFT}=\frac{l_s}{8}$}
  \label{fig:win_size_1_8}
\end{subfigure}
\begin{subfigure}{.33\textwidth}
  \centering
  % include second image
  \includegraphics[width=\linewidth]{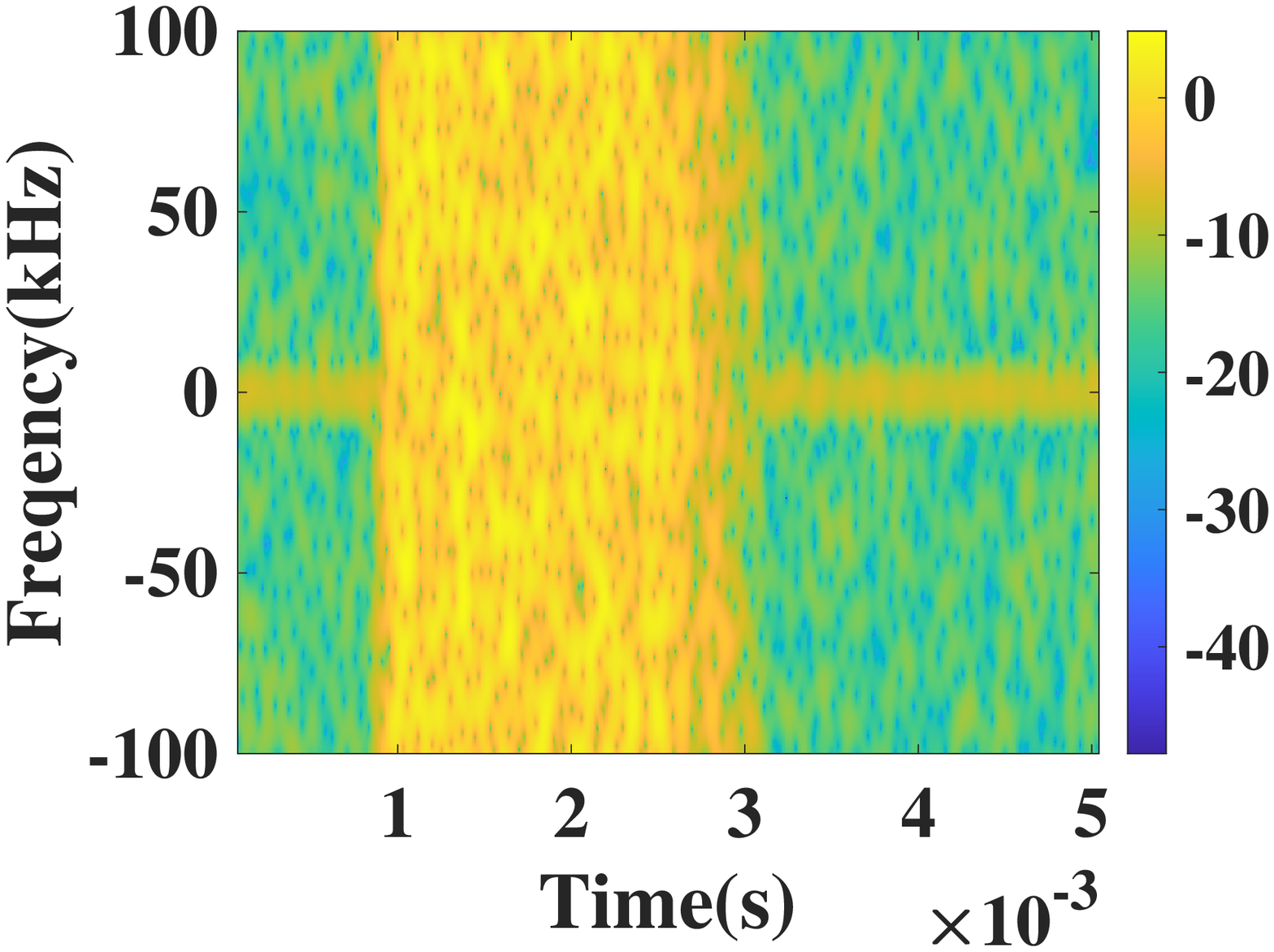}
  \caption{$l_{STFT}=\frac{l_s}{32}$}
  \label{fig:win_size_1_32}
\end{subfigure}
\caption{Different STFT Window size}
\label{fig:dif_win_size}
\vspace{-4mm}
\end{figure*}

\begin{figure}[!htb]
\centering
\includegraphics[width=0.45\textwidth]{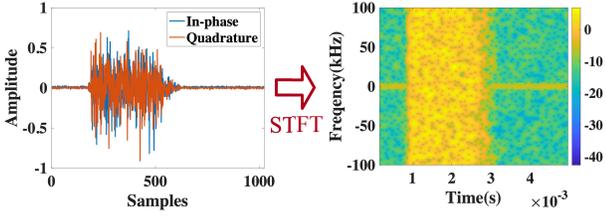}
\caption{An example of performing STFT to dechirped symbol}
\label{fig:STFT_results}
\vspace{-4mm}
\end{figure}

\subsubsection{\textbf{Frequency-Time Analyse}}

The LoRa PHY layer applies the CSS modulation method, and the frequency linearly increases with time. Traditional LoRa performs FFT to the entire symbol-length chip sequence, which extracts the frequency bin but abandons the information in the time domain. STFT provides the time-localized frequency information for situations in which frequency components of a signal vary over time, whereas the standard Fourier transform provides the frequency information averaged over the entire signal time interval. The main idea of STFT is to intercept the signal and Fourier transforms the intercepted signal to get the time spectrum of the signal.

 \begin{equation}\label{stft}
z_s(\tau,m)=\sum_{n=0}^{N-1}x(n)w(n-\tau)e^{\frac{-i2\pi mn}{N}}
\end{equation}

Eq.(\ref{stft}) shows the basic form of STFT. Here $w(n-\tau)$ is the Hann window function. with $w(n-\tau)=0,|n-\tau|>\frac{l_{STFT}}{2}$  STFT provides a coarse-grained estimate of chirp frequency, which cannot precisely identify the frequency track of LoRa chirp. 

To solve the problem that clean chips in interference signals are hard to be selected for recovery, PSR separates the clean chips in the corrupted symbol by exploiting the frequency-domain and time-domain features. We find that the frequency of the LoRa chirp changes continuously in the STFT window, as there is no CTI in the window. Whereas for the condition of low SNR, as the frequency feature vanishes with the STFT window of LoRa chirp. Since the bandwidth of overlapping CTI is generally larger than that of LoRa signals, the frequency feature of CTI is similar to the white noise which is indistinguishable from LoRa signals in low SNR. Based on this observation, we first do a dot multiplication of the LoRa symbol and downchirp. The frequency of the LoRa signal after dechirped does not change linearly with time, but a fixed value. As Fig. \ref{fig:STFT_results} shows, the operation of STFT collects the number of window size dechirped LoRa chips to get a gain for low SNR condition.

STFT is limited by Heisenberg's uncertainty principle and cannot achieve good energy gathering performance in both time domains and frequency domains. Figure\ref{fig:dif_win_size} present the resulting spectrograms of the same signal with different size of STFT window. Comparing Figure \ref{fig:win_size_1_8} and Figure \ref{fig:win_size_1_32}, we see that time series in the spectrogram become shorter, because larger window size contain less sliding space. With Figure \ref{fig:win_size_1_8} and Figure \ref{fig:win_size_1_32}, we see that a larger window achieves better processing gain, leading to more accurate frequency results. Specifically, the corresponding frequency point has more spectral components, and in the time-frequency diagram, the corresponding bright line is brighter. But a larger window means that it is easier to account for interference. If there is interference in the calculated chip sequence (Figure \ref{fig:win_size_1_2}), there will be no concentrated spectral components in the STFT results, and the spectral components are closer to an even distribution. There will be not only one bright line in the time-frequency spectrograms. On the contrary, a smaller window process the chip sequence fine-grained, but gets less processing gain, which will produce a wide main-lobe. A window that is too large or too small both causes the loss of bright lines in the time-frequency spectrogram. Therefore, we selected six windows of different sizes to do STFT processing on the interfered symbol. The size of these six windows is not fixed, but a value that is positively correlated with the 
SF. Six windows of different sizes sequentially perform independent STFT operations on the dechirped signal sequence. The result of the larger window provides a lower limit of the interference criterion, and the result of the smaller window is responsible for extracting as many clean chips as possible.

\subsubsection{\textbf{Max Pooling}}

 \begin{figure}[!htb]
 \centering
 \includegraphics[width=0.48\textwidth]{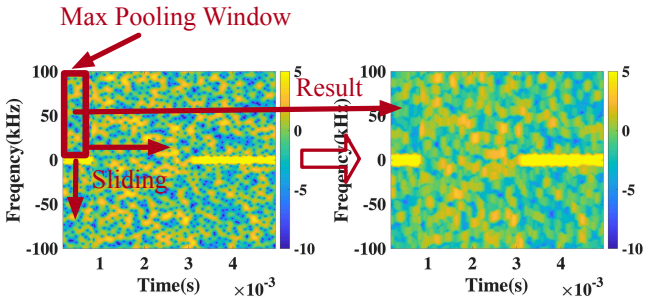}
 \caption{Max Pooling}
 \label{fig:maxpooling}
\vspace{-4mm}
 \end{figure}

Since the size of the STFT window we used is smaller than the number of chips in the symbol, this causes spectrum leakage which leads the bright lines to not stand out. This feature is more obvious in a low SNR condition. In the module of error detection, we do not rely on the information in the spectrogram for symbol demodulation, but extract more clean chips through the characteristic bright line in the spectrogram. In order to reflect this bright line more clearly, we use a special STFT window and an operation of Max Pooling to process it. Specifically, we first consider using the Hanning window to concentrate the spectral components in the middle when performing STFT. Considering the existence of noise and less processing gain, the maximum value in each time slot of the STFT result may not be the accurate signal frequency bin, but distributed near the real signal frequency bin. Our main purpose is to highlight the bright lines in the time-frequency spectrograms, and to tolerate the broadening of the bright lines of the spectrum. To this end, we use the Max Pooling window to perform convolution processing on the frequency domain of the signal STFT result to further improve the feature.

 \begin{equation}\label{MaxPooling}
Y_{pool}[\tau,m]=argmax_{m-\frac{l_{pool}}{2}\leq p \leq m+\frac{l_{pool}}{2}}z_s(\tau,p)
 \end{equation}

Eq.(\ref{MaxPooling}) shows the basic form of Max Pooling. Through the operation of  Max Pooling, the value of a certain point in the time-frequency spectrograms is not only related to its own value, but the maximum value of nearby frequency bins within the same time slot. In order to better match the STFT window, the size of the Max Pooling window is positively related to the size of the main lobe of the Hanning window

The bright line in the LoRa spectrogram we mentioned earlier, its meaning is the corresponding spectral component after LoRa symbol dechirped. We can see from the frequency spectrum that the part of the interfered signal does not have this bright line. This is because the interfered signal is a broadBand signal compared to the LoRa signal. After the interfered signal is dechirped, it will be approximately flat in the entire spectrum. Compared with the LoRa signal under low SNR, the spectrum component of the interference signal after Dechirped is larger. If the existing time-frequency sequence is not processed, our operation of taking the largest bright line in horizontal based on the spectral components is easily affected by interference signals. And the large spectral components of the interference signal cause our positioning of the bright line to be almost random. In order to better distinguish interfered and clean chips, we calculate the ratio of the value of each frequency component in each slot in the spectrogram to the average value of the frequency components in the slot. After this operation, the original interference signal with higher spectral components is suppressed, and the LoRa signal is not affected.

 \begin{equation}\label{ratio}
X_{ratio}[\tau,m]=\frac{Y_{pool}[\tau,m]}{\sum_{m=0}^{N} Y_{pool}[\tau,m]}
 \end{equation}

Eq.(\ref{ratio}) shows the form of the current frequency component to the average ratio.  Using the difference between the LoRa signal and the interference signal at this value, we can distinguish the interference signal from the received signal. We scan the maximum value of the $X_{ratio}[\tau,m]$ sum corresponding to each frequency bin to locate this bright line. Since the sizes of the STFT windows we used are different, and the results of processing gain are also different, we normalize the calculated $X_{ratio}[\tau,m]$ value.

 \begin{equation}\label{norm} 
X_{norm}[\tau,m]=\frac{X_{ratio}[\tau,m]}{l_{STFT}}
 \end{equation}
 
Eq.(\ref{norm}) shows the form of normalized current frequency component to average ratio. We save $X_{norm}[\tau,m]$ to help the fine-grained symbol detection process. STFT windows of different sizes generate multiple $X_{norm}[\tau,m]$ time-frequency sequences, we correspond them one by one. By comparing the values of $X_{norm}[\tau,m]$ sequences, we realize interference discrimination. Specifically, the larger the value in $X_{norm}[\tau,m]$, the higher the ratio of the spectral components to the entire spectrum. That is, the main features of the LoRa signal are obvious, and we judge the chips corresponding to this value as interference free.  
 
 \begin{figure}[!htb]
 \centering
 \includegraphics[width=0.48\textwidth]{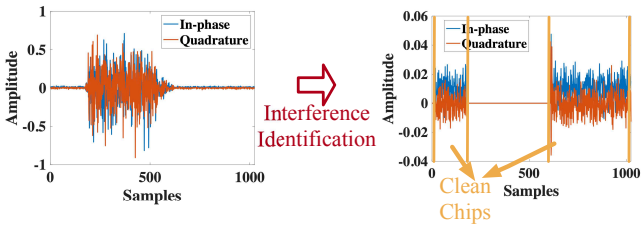}
 \caption{Interference Identification}
 \label{fig:interference_identification}
\vspace{-4mm}
 \end{figure}

\subsection{Fine-grained symbol detection} 
In this section, we present our design for accurately identifying clean LoRa chips, followed by the symbol recovery.

\subsubsection{\textbf{Chip level Interference identification}}

After the operation mentioned before LoRa has unique characteristics in the time-frequency sequence. Specifically, for the value in the same time slot, the LoRa signal has a peak on one frequency bin, and the values of other frequency bin are similar and low. We expect to extract fine-grained clean chip sequences from the interfered symbol. 
  
Each value in the time-frequency sequence represents the spectral component value of multiple chips corresponding to the sliding window. A larger value means that more clean chips are calculated, corresponding to the LoRa signal that is not interfered. As mentioned before, the spectral component value is related to the STFT window size, that the larger size, the higher the gain. Therefore, We first select the larger time-frequency sequence value in the large-size window. The large time-frequency value is mapped to multiple clean chips in the symbol which is utilized for recovery. Since a LoRa symbol could be interfered by multiple CTI packets, the clean chip sequence may be cut into many segments. The clean chips of a maximum time-frequency value mapping are not sufficient to realize symbol recovery. We calculated a recovery threshold for the number of clean chips, which is closely related to SNR and SF. Compared with the threshold, if the number of clean chips is insufficient, we traverse the next largest time-frequency sequence value and extract the mapped new clean chips and merge them with the previous chips. Iterate this process until the number of total selected clean chips is larger than the recovery threshold, and then perform correlation demodulation for recovery.

Figure \ref{fig:interference_identification} demonstrates an example of identifying clean LoRa chips. In order to better show, we remove the interfering part of the signal in the right figure of Figure \ref{fig:interference_identification}. Specifically, the LoRa chips
within [1,176] and [634,1024] range  are not interferred by wireless interference. By examining the $X_{norm}[\tau,m]$ matrix, our design manages to accurately identify the parts free of interference, as demonstrated in Figure \ref{fig:interference_identification}.

\subsubsection{\textbf{Correlation demodulation}}

In the previous section, we introduce how to coarsely locate the transmitted LoRa symbol. In this section, we need to identify clean LoRa chips free of interference and then perform partial symbol recovery.

After detecting clean LoRa chips, we utilize clean chips to realize partial symbol recovery. This is achieved by computing the dot product between the clean LoRa chips and the standard downchirp signal. The downchip is the conjugate signal of the upchirp, and its frequency linearly decreases with time.
We calculate the correlation between the frequency of the multiplied signal and the corresponding frequency of different symbols to achieve demodulation. Specifically, we compute FFT processed signal to transform it to the frequency domain and choose the bin corresponding to the frequency with the highest correlation as the demodulation result.

\begin{figure}[!htb]
 \centering
 \includegraphics[width=0.48\textwidth]{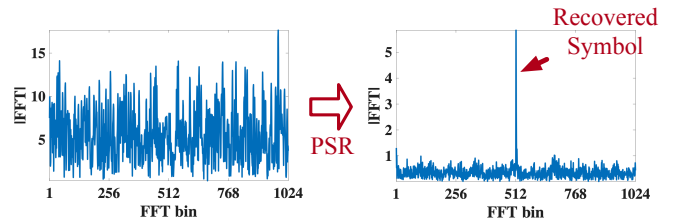}
 \caption{Recovery Result}
 \label{fig:Demodulation}
\vspace{-4mm}
 \end{figure}

Figure \ref{fig:Demodulation} (a) shows the FFT result for a corrupted LoRa symbol. It is clear that the LoRa signal is destroyed by interference, since there is not a dominating frequency component.
After our PSR, Figure \ref{fig:Demodulation} (b) demonstrates the FFT result of symbol recovery. There is a dominating frequency component at 513 with much larger energy than other frequency components. Figure \ref{fig:Demodulation} means that the frequency bin has the highest correlation at 513.

By identifying clean LoRa chips and removing interference from demodulation, our PSR manages to correctly recover the transmitted symbol.

\section{Discussion}

\subsection{General Applicability}
This paper focuses on enhancing the CTI protection for the 2.4 GHz as a proof of concept, while its design principles could be generally adopted to other technologies in the unlicensed spectrum. 
This is because of the inherent nature of the long symbol duration for LPWAN communication, which is targeting long-range communication. In contrast, CTI has a limited duration and only corrupts a portion of one LPWAN symbol. As a result, by identifying and utilizing the correct signals at the PHY layer, our design could further enhance the decoding performance.
For example, our design could be directly applied for enhancing the LoRa in the 915 MHz, which has a spectrum overlapping with emerging 802.11 ah. In addition, our technique also offers protection against different CTI sources, e.g., ZigBee and Bluetooth, since it does not require specific knowledge of the CTI.

\subsection{Header Protection}
Our design elaborately recovers the corrupted LoRa symbols at the PHY layer, while it could also be adopted for header protection. In current LoRa packets, packet headers contain the critical information and are protected via coding and cyclic redundancy check (CRC). For headers, our design could be applied to provide additional protection.

\subsection{Complexity}
Since our PSR requires accessing the PHY level signal, it requires modification of the PHY layer at LoRa receivers. However, our technique is lightweight by only examining the features of the LoRa signal. Different from our approach, 
existing interference cancellation techniques \cite{yubo2013zimo,zheng2014zisense,gollakota2008zigzag} require a very high sampling rate (e.g., 20 MHz) and the implementation of the demodulation procedures of the interference protocols, e.g., WiFi.
This leads to significant implementation complexity, given the increasingly dense and heterogeneous IoT environment.

We give a specific derivation process to explain the complexity of PSR. There are $N$ chips in a symbol, and the width of each STFT window is $l_{STFT}$. The complexity of performing an FFT on n chips is $O(N*log N)$. For a interfered symbol, we slide the STFT window from the beginning to the end, and the complexity of  the two-dimensional spectrogram is $O((N-l_{STFT})*N*log N)$. In order to get different grained clean chip blocks, we utilize $m$ STFT windows of different sizes to slide on the interfered symbol. In summary, the complexity of our design is $O(m*(N-l_{STFT})*N*log N)$. Generally, $l_{STFT},m$ are much smaller than $N$, and the final complexity of our design is $O(N^2*log N)$. Compared with the existing LoRa demodulation design, PSR increases the complexity of $N$ times, which is tolerable on existing the COT LoRa nodes.

\begin{figure}[!htbp]
\centering
\includegraphics[width=0.45\textwidth]{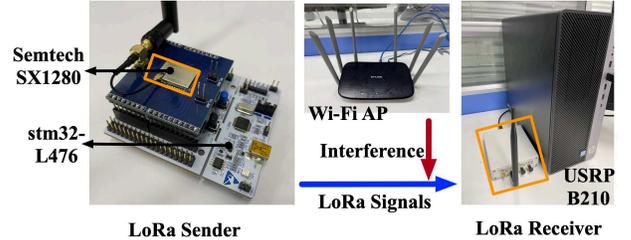}
\caption{Evaluation setup}
\label{fig:evaluation_setup}
 \vspace{-6mm}
\end{figure}

\begin{figure*}[!htb]
\begin{subfigure}{.33\textwidth}
\centering
  % include first image
  \includegraphics[width=\linewidth]{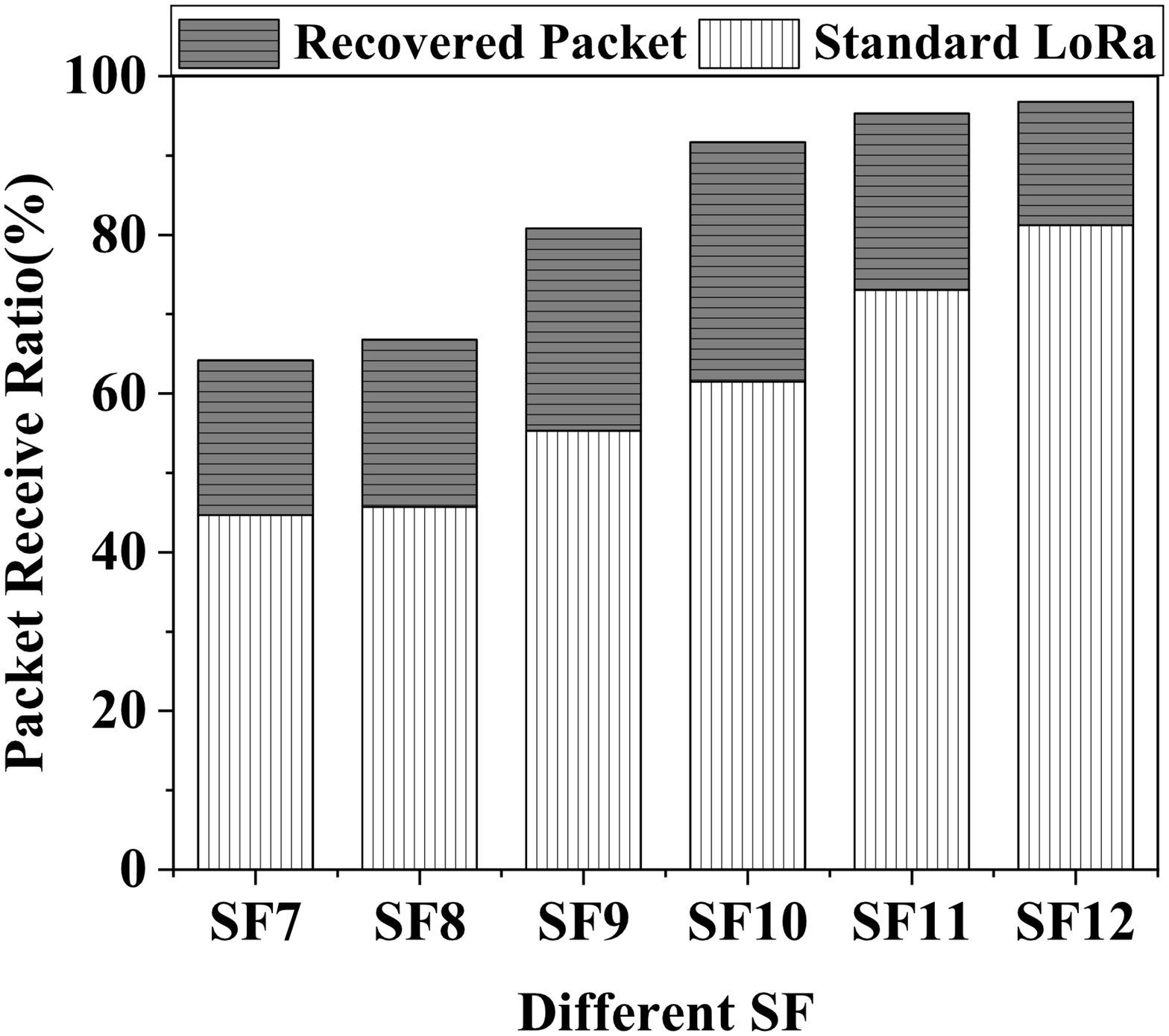}
  \caption{PRR in different SF (Low Interference)}
  \label{fig:prr_dif_sf}
\end{subfigure}
\begin{subfigure}{.33\textwidth}
\centering
  % include first image
  \includegraphics[width=\linewidth]{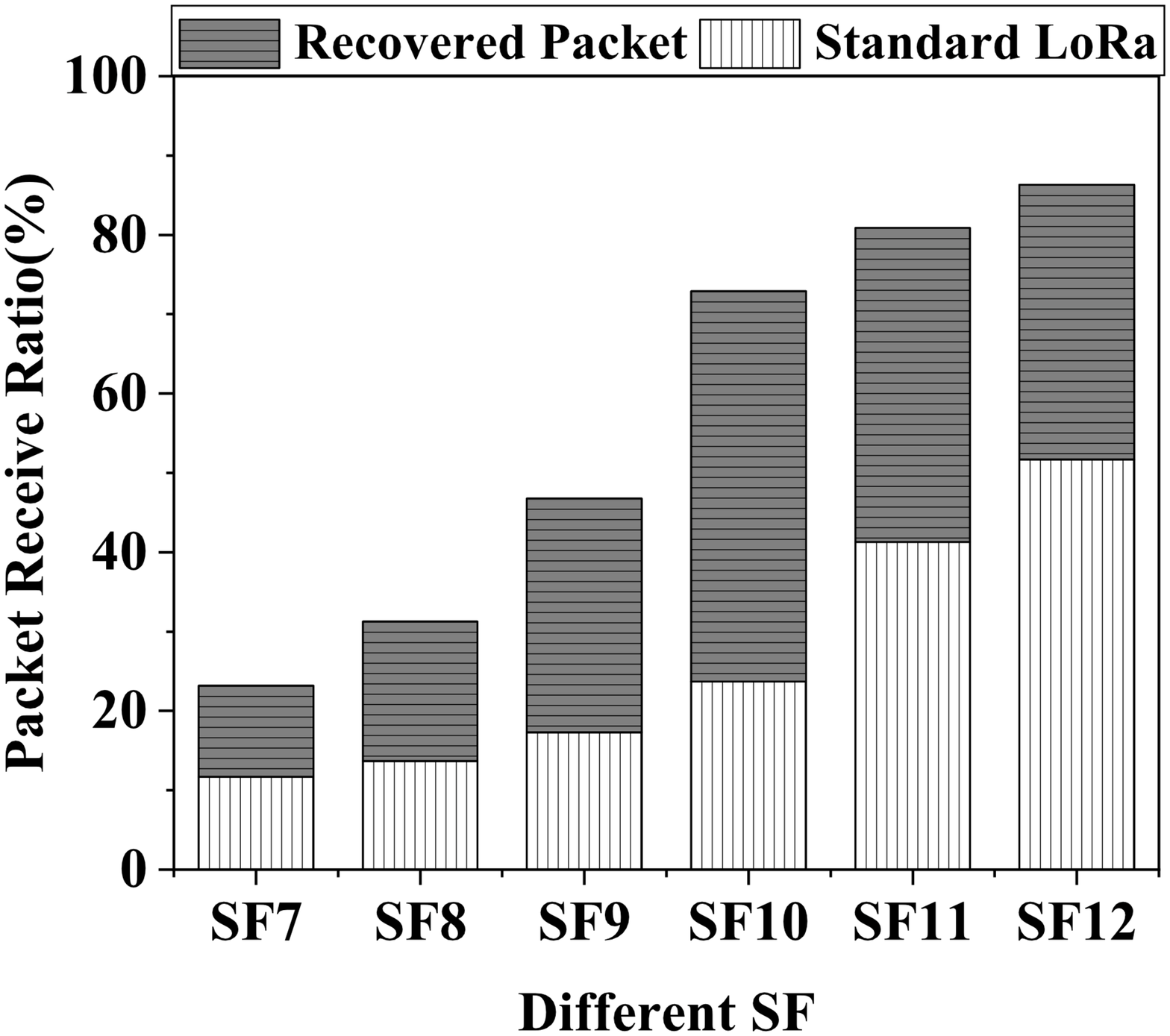}
  \caption{PRR in different SF (Mid Interference)}
  \label{fig:prr_dif_sf_mid}
\end{subfigure}
\begin{subfigure}{.33\textwidth}
\centering
  % include first image
  \includegraphics[width=\linewidth]{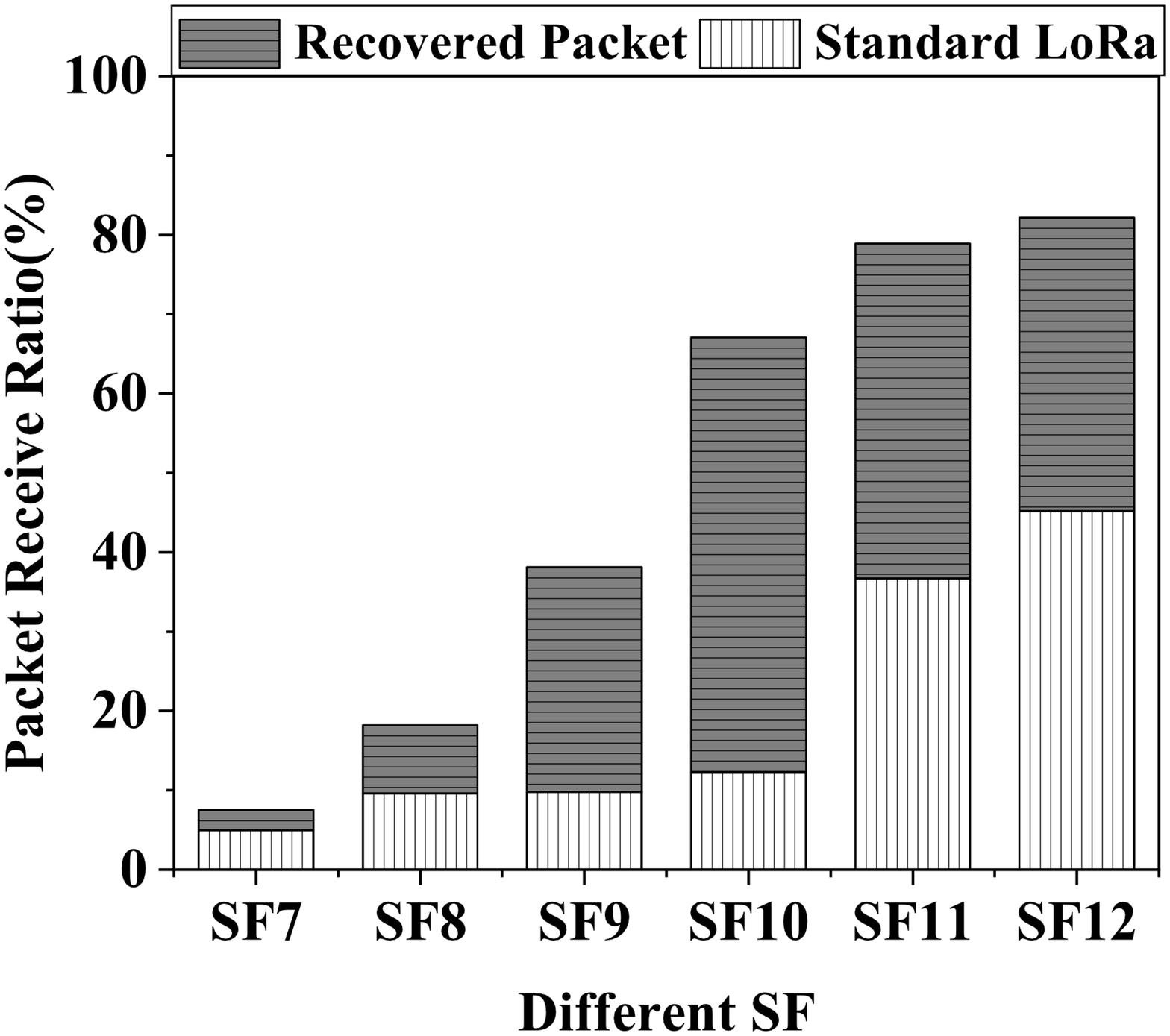}
  \caption{PRR in different SF (High Interference)}
  \label{fig:prr_dif_sf_high}
\end{subfigure}
\caption{Packet recovery performance under different interference}
\label{fig:prr_dif_traffic}
 \vspace{-4mm}
\end{figure*}

\begin{figure*}[!htb]
\begin{subfigure}{.33\textwidth}
\centering
  % include first image
  \includegraphics[width=\linewidth]{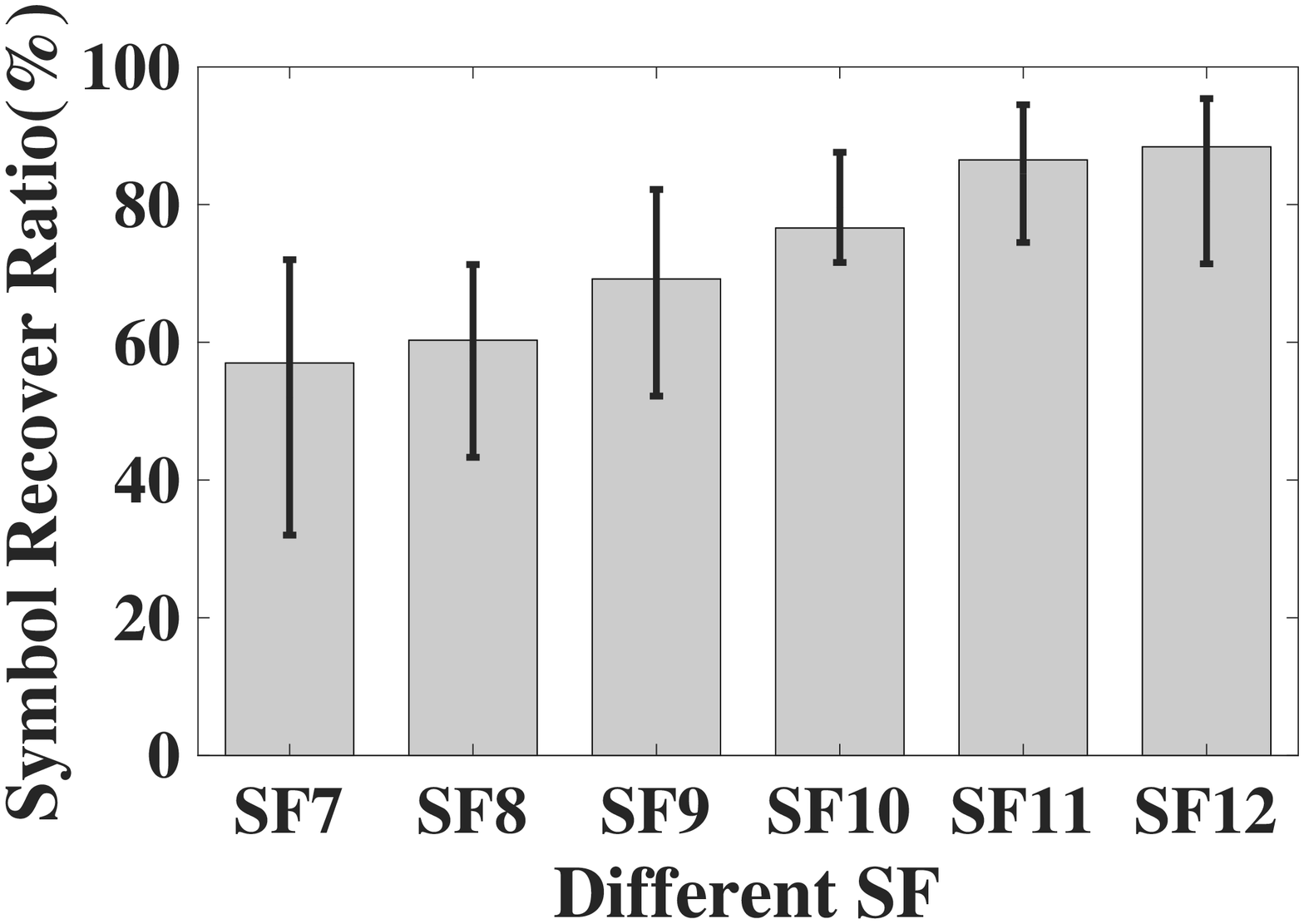}
  \caption{SRR in different SF (Low Interference)}
  \label{fig:srr_dif_sf}
\end{subfigure}
\begin{subfigure}{.33\textwidth}
\centering
  % include first image
  \includegraphics[width=\linewidth]{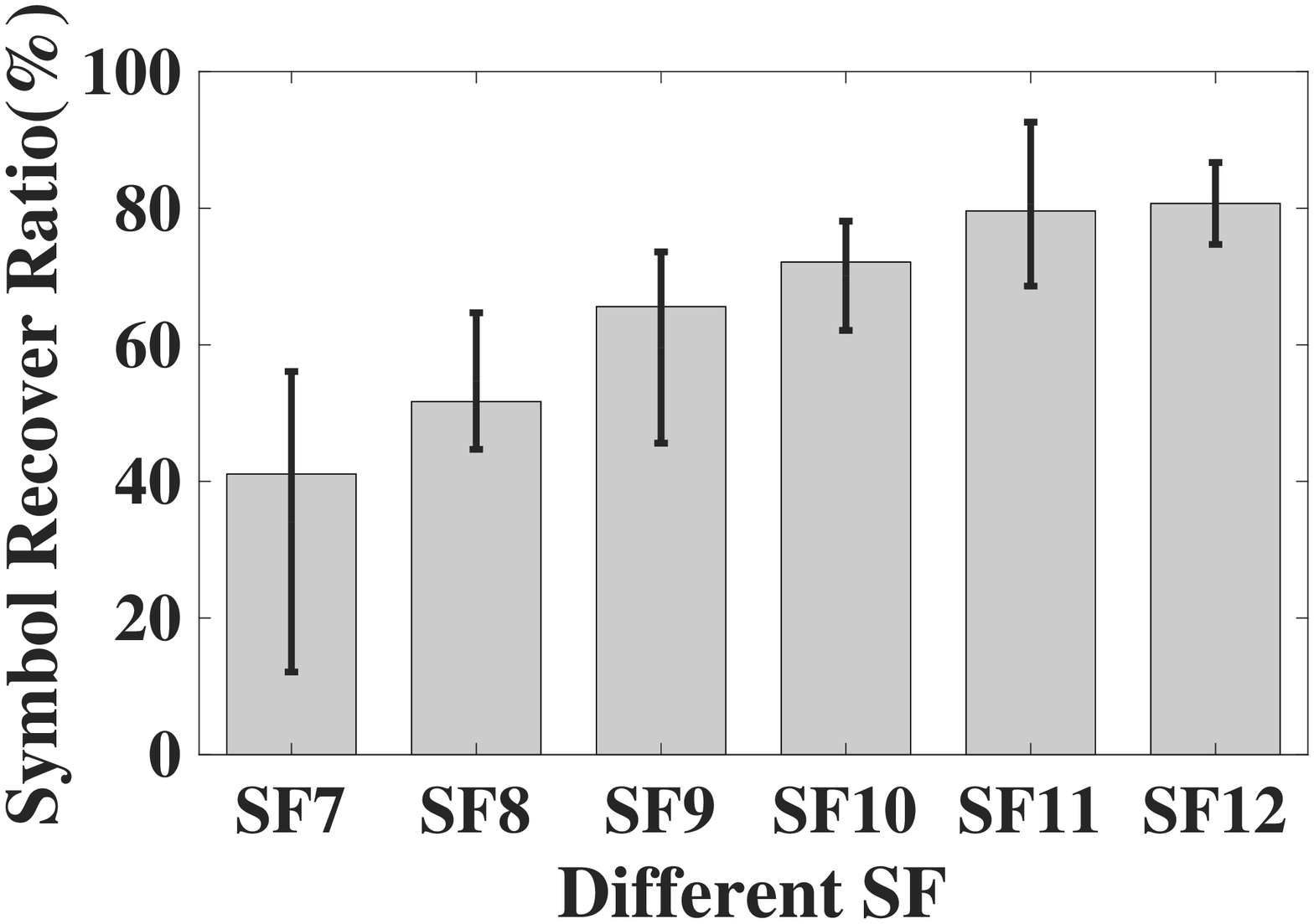}
  \caption{SRR in different SF (Mid Interference)}
  \label{fig:srr_dif_sf_mid}
\end{subfigure}
\begin{subfigure}{.33\textwidth}
\centering
  % include first image
  \includegraphics[width=\linewidth]{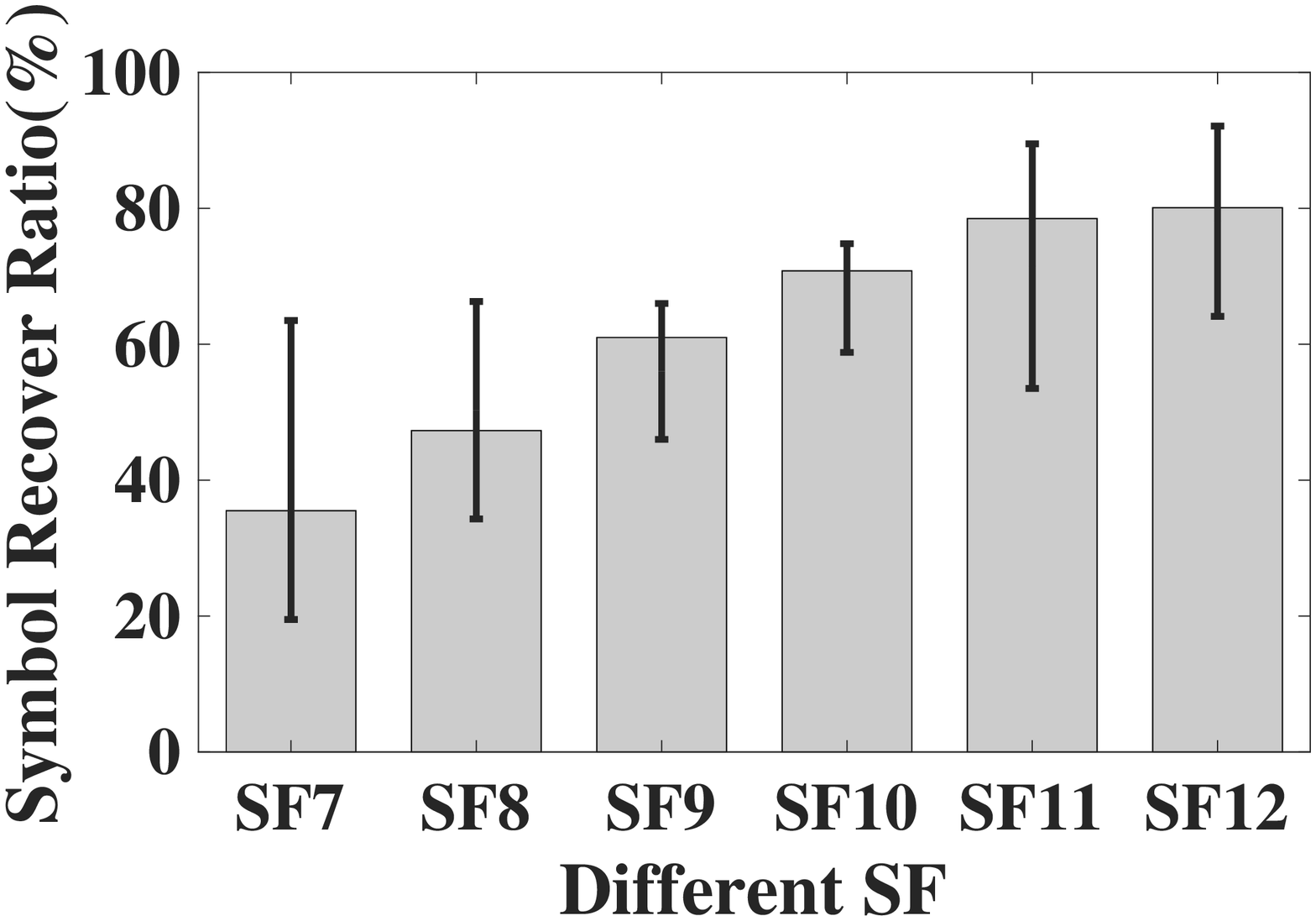}
  \caption{SRR in different SF (High Interference)}
  \label{fig:srr_dif_sf_high}
\end{subfigure}
\caption{Symbol recovery performance under different interference}
\label{fig:srr}
 \vspace{-4mm}
\end{figure*}

\section{Evaluation}
This section presents the evaluation of PSR with real testbeds.
Figure \ref{fig:evaluation_setup} presents our customized testbed, which includes SX1280 LoRa transceiver chip and stm32-L476 controller, for transmitting LoRa packets.
As for the receiver side, we utilize USRP B210 to capture the wireless signal at the normal LoRa sampling rate and implement our 
PSR based on the collected samples.
To evaluate the resilience of PSR, we test it under the typical wireless interference at 2.4GHz, e.g., WiFi, ZigBee, and Bluetooth.
Since the WiFi is the dominating interference in the 2.4GHz, our evaluation focuses on the interference due to WiFi traffic, while Section analyzes PSR's performance under ZigBee and Bluetooth.

We first present the overall protection performance (packet reception ratio), followed by detailed experiments under different settings for examining our design.

\subsection{Performance Overview}\label{evaluation_overview}
Figure \ref{fig:prr_dif_traffic} demonstrates the packet reception ratio (PRR) of standard LoRa and our design under different SF parameters and different volumes of Wi-Fi interference. The standard LoRa uses Hamming code with code rates 4/8, interleaving code and gray code to improve the performance for short bursty interference at low SNR.
For the purpose of controlled experiments, we utilize the commodity WiFi cards to generate WiFi traffic at different volumes.
Specifically, we connect different numbers of wireless devices to the Wi-Fi AP to generate different traffic scenarios. 
For the low WiFi interference, there is one WiFi device connected to the WiFi AP, with an average of 350 WiFi packets per second.
For the medium WiFi interference, 3 WiFi devices are connected to the WiFi AP, leading to 1,500 packets/s.
For the high WiFi interference, there is 5 working WiFi in total, while the number of WiFi packets is 2,600 packets/s.
In each experiment, we transmit more than 2,000 LoRa packets at the length of 100 bytes in every SF.
Considering the typical scenarios and long-range feature of LoRa,
the signal-to-noise ratio (SNR) of LoRa is usually low \cite{zhou2019design}.
Without loss of generality, the SNR of LoRa is -10dB in our experiments, similar to the SNR reported in \cite{zhou2019design}.

Under low Wi-Fi interference, LoRa still suffers from packet losses. For example, when SF = 10, the packet loss ratio is 38.5\%.
Via our design, we successfully improve the packet reception ratio from 61.5\% to 91.7\%.
With the increase of WiFi traffic amount, standard LoRa suffers from more packet corruption. 
For example, under high Wi-Fi interference and an SF value of 10, standard LoRa only has a reception ratio of 12.2\%. 
In contrast, our design manages to boost the packet reception ratio to 67.1\%, 5.5 times the standard LoRa.

To offer the analysis on the protection of PSR, we also analyze the symbol recovery ratio (SRR) for the previous experiment. The symbol recovery ratios are demonstrated in Figure \ref{fig:srr_dif_sf}. 
When the Wi-Fi interference is low, our design demonstrates good reliability - more than 57\% of the corrupted symbols could be recovered when the interference is low. 
With the increase of Wi-Fi traffic intensity, the recovery ratio slightly decreases, from 57.1\% to 35.5\%.

This series of experiments demonstrate the good reliability of PSR, which is achieved by utilizing the unique features of partially correct signals in LoRa communication.

\begin{figure}[!htbp]
\centering
\includegraphics[width=0.37\textwidth]{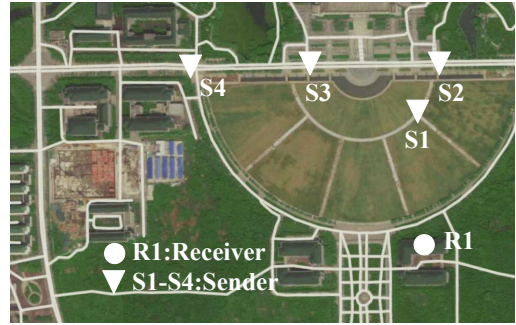}
\caption{Outdoor experiment scene}
\label{fig:outdoor_setting}
 \vspace{-4mm}
\end{figure}

\begin{figure}[!htb]
\centering
  % include first image
  \includegraphics[width=0.74\linewidth]{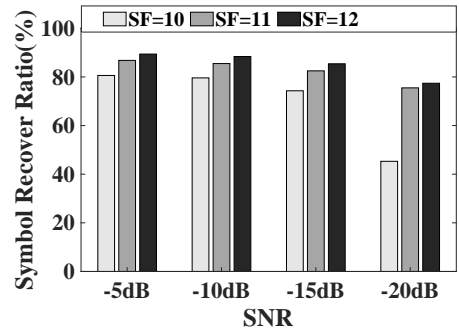}
  \caption{Symbol recover ratio under different SNR}
  \label{fig:evaluation_symbol_sf}
   \vspace{-6mm}
\end{figure}

\subsection{SNR Impacts}
We also evaluate the reliability of PSR under different SNR. 
Our experiments are conducted on a university campus. We set up four different sender locations from near to far corresponding to S1 to S4 in the Figure \ref{fig:outdoor_setting}, and they are 150m, 200m, 350m, 500m away from the receiver. Setting different sending locations is to evaluate the system performance under different channel quality, and their corresponding SNR values are -5dB, -10dB, -15dB and -20dB.

Figure \ref{fig:evaluation_symbol_sf} shows the result. 
Even under a very low SNR of -20 dB, our design manages to maintain a symbol recovery ratio of 77.4\% for SF=11.
This is because PSR elaborately considers the inherent low SNR by analyzing the frequency-time pattern with STFT at different window sizes.

\begin{figure}[!htbp]
\centering
\includegraphics[width=0.37\textwidth]{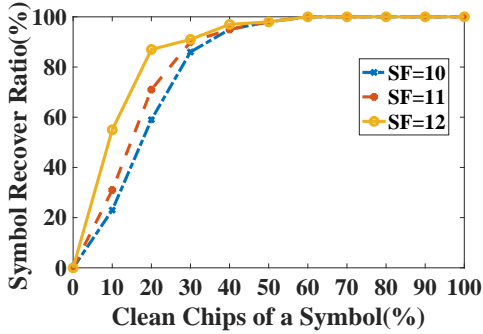}
\caption{Symbol recovery performance with different clean chips number}
\label{fig:srr_dif_part}
 \vspace{-4mm}
\end{figure}

\subsection{Impacts of Interference Duration}
To provide a detailed examination of our system, we collect 1,000 corrupted LoRa symbols (with SF10, SF11 and SF12). Then we measure the duration of chips that are free of
Wi-Fi interference and utilize our design to recover them. 
Figure \ref{fig:srr_dif_part} depicts the error correction performance with interference duration. It is clear that our design works better when symbols have less percentage of interference. For example, when the corrupted symbol has more than 40\% of clean chips, our design rarely fails. 
When the percentage of clean chips drops under 20\%, the recovery suffers from failures due to the limited clean chips within a LoRa symbol. 

Figure \ref{fig:srr_dif_part} depicts the recovery performance for SF11 and SF12 too. It is clear that LoRa symbols with larger SF have higher chances of successful recovery. For example, the recovery probability for SF12 is 87\%, while it is only 59\% for SF10. This is because LoRa symbols with higher SF have a longer duration, thus offering better resilience.

\begin{figure}[!htb]
\centering
  % include first image
  \includegraphics[width=0.74\linewidth]{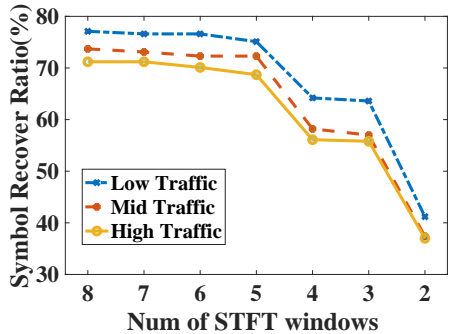}
  \caption{Impacts of the number of STFT windows }
  \label{fig:evaluation_window}
   \vspace{-4mm}
\end{figure}

\subsection{Impacts of STFT Windows}
Figure \ref{fig:evaluation_window} examines the number of STFT windows on the symbol recovery ratio.
With an STFT window size of 6, our system achieves a good symbol recovery ratio. When the number of STFT window decreases, the recovery ratio starts to drop. 
In a scene of dense interference, a LoRa symbol could be interfered by multiple interference packets. The number of windows decreases, if the STFT window is large, the interference could be included in each sliding STFT window. If the window is small, there is no spectrum concentration effect, making it hard to extract a sufficient number of clean chips for recovery. Different interference conditions have different requirements for the size of the STFT window. Therefore, we utilize multiple sliding windows of different sizes to perform STFT processing on the same signal.
As shown in Figure \ref{fig:evaluation_window}, when the number of STFT windows increases from 6, our system has stable performance.
Since more STFT windows lead to more FFT computations, we choose the number of STFT windows to be 6 for achieving the balance between performance and complexity.

\begin{figure}[!htb]
\begin{subfigure}{.24\textwidth}
\centering
  % include first image
  \includegraphics[width=\linewidth]{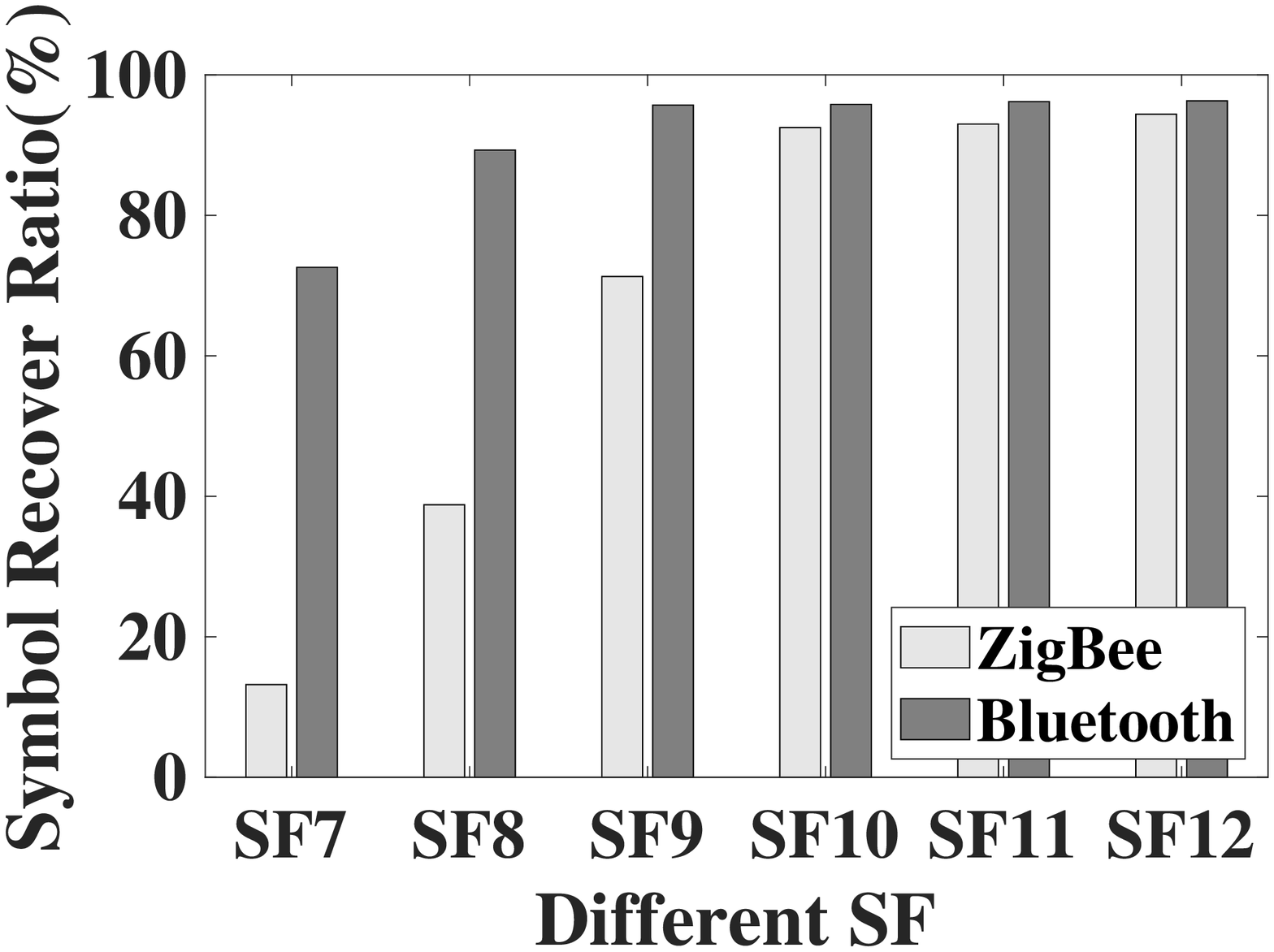}
  \caption{Symbol Recover Ratio}
  \label{fig:srr_dif_sf}
\end{subfigure}
\begin{subfigure}{.24\textwidth}
\centering
  % include first image
  \includegraphics[width=\linewidth]{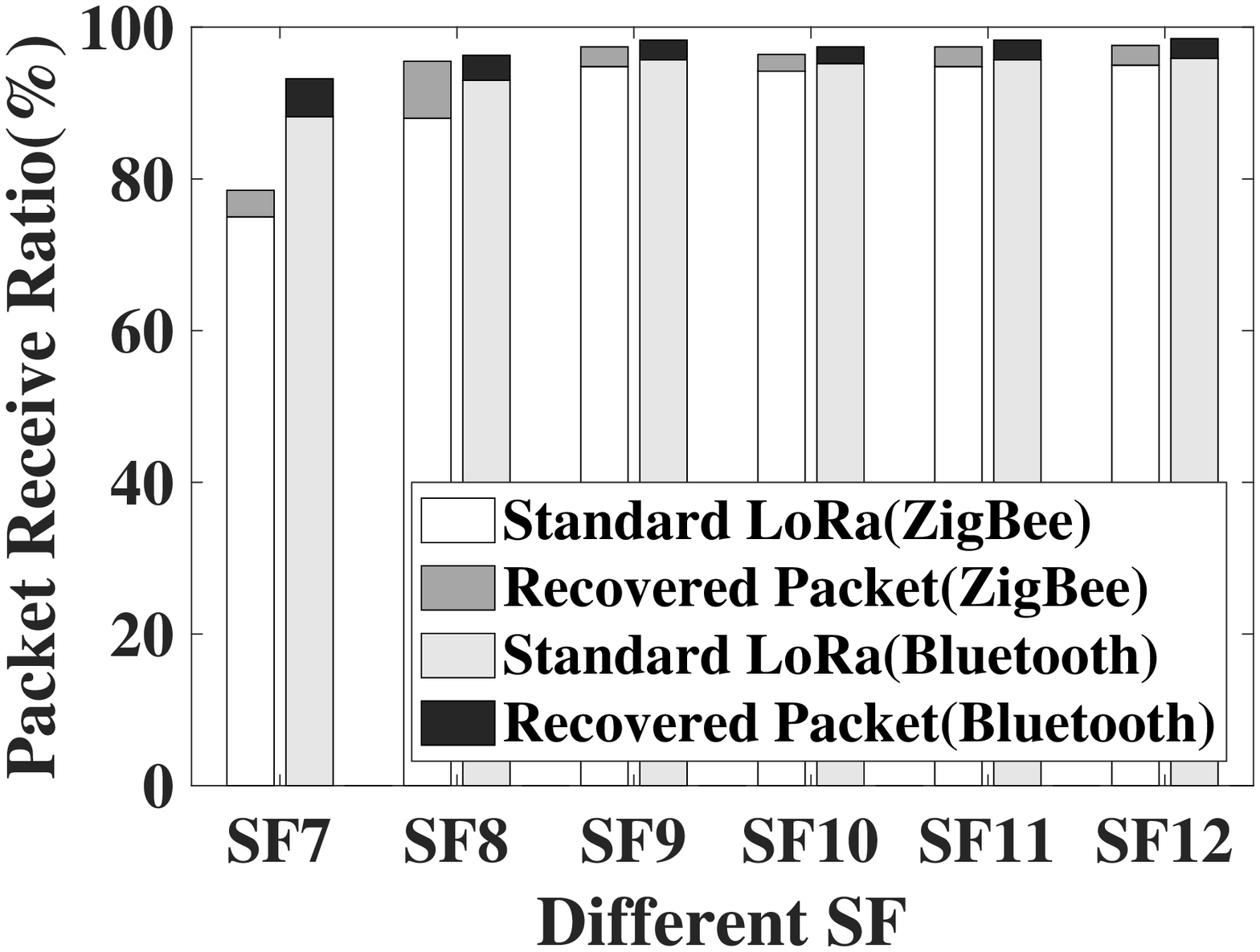}
  \caption{Packet Receive Ratio}
  \label{fig:prr_dif_sf}
\end{subfigure}
\caption{SRR and PRR under different interference}
\label{fig:evaluation_prr_zigbee_bluetooth}
\vspace{-4mm}
\end{figure}

\subsection{General Protection}
Our system is a general design, which is effective in recovering the LoRa symbol under the interference of ZigBee and Bluetooth in the overlapping frequency. Figure \ref{fig:evaluation_zigbee_bluetooth} shows the relationship between symbol recover ratio and SF, while the main interference is ZigBee or Bluetooth in this experiment. The results show that the symbol recover ratio rises from 13.2\% in SF7 to 94.4\% in SF12 under ZigBee interference. And the symbol recover ratio maintains a good performance in different SF which are all higher than 72.6\%. Compared with Wi-Fi and Bluetooth interference, 
ZigBee shows more destruction, while ZigBee has a longer packet duration which destroys the entire LoRa symbol in small SF(7, 8 and 9) causing there is not enough clean chip in an interfered symbol to recover.

\begin{figure}[!htb]
\centering
 % include first image
 \includegraphics[width=0.8\linewidth]{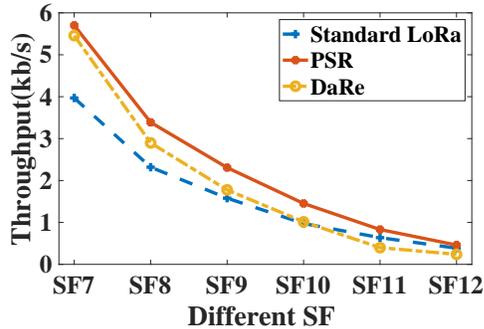}
 \caption{Throughput comparison}
 \label{fig:evaluation_throughput}
 \vspace{-4mm}
\end{figure}
\subsection{Throughput}

This section shows the LoRa throughput in different SF under Wi-Fi traffic.
Suffering from interference, LoRa packets may have corrupted symbols and thus fail the CRC check. 
The packet loss directly limits the LoRa throughput performance.
We compare the standard LoRa, the DaRe \cite{marcelis2020dare} and our PSR to show the throughput of the system under interference.

Figure \ref{fig:evaluation_throughput} demonstrates the LoRa throughput under different SF.
When parameter SF is small, our throughput is slightly larger than DaRe, which is 1.05 times DaRe in SF7. 
The large SF could be extracted more clean chips in one symbol, our PSR gets a higher process gain with increasing of SF. PSR reaches a throughput of 0.46 kb/s, which is 1.2 times standard LoRa and 1.9 times DaRe.
To improve robustness, DaRe utilizes redundant information calculated from previous data packets to encode packets. Because of the excessive extra redundancy added to the LoRa payload, it actually leads to a decrease in throughput compared to the standard LoRa Dare. On the other hand, DaRe needs to wait for the next successful transmission for recovery, and it leads to a substantial increase in system latency. A LoRa packet with 100 symbol in SF12 has a Time-on-Air of 2000 ms, which is the lower limit of the system latency.
In contrast, compared with other designs our PSR maintains a relatively stable throughput service under dense CTI in different SF.

\section{Related Works}
\par Recent years have witnessed the proliferation of Internet of Things (IoT) applications, such as smart agriculture\cite{ilie2016monitoring}, smart house \cite{de2012state}, environmental monitoring, building automation\cite{liang2016systematically}  military field \cite{furtak2016security} and smart cities \cite{batty2012smart}\cite{chourabi2012understanding}. These IoT application great enrich our daily life \cite{saquib2017blindar} \cite{neto2016aot}, calling for effective wireless protocols to connect deployed IoT devices at long range with low power.
\par Low-power wide area network(LPWAN) has been recently introduced to offer connectivity to the low-cost devices distributed over very large geographical areas \cite{liando2019known}. Because of this reason, LPWAN has a limited transmission speed, e.g., from 0.3 kbps to 50 kbps \cite{LoRawan_ref}. To enable these goals, researchers have introduced various protocols, such as LoRaWAN \cite{LoRawan_ref}, SigFox \cite{sigfox_ref}, NB-IoT \cite{adhikary2016performance} and LTE-M \cite{lte_m_ref}. Among these LPWAN techniques, LoRa works in the unlicensed spectrum for removing the cost of spectrum access, while it has been widely adopted across regions for offering connectivity \cite{mekki2019comparative} \cite{mekki2018overview}. \cite{petajajarvi2017evaluation} evaluates the performance of LoRa LPWAN.
\par There has been extensive literatures on LoRa, while they mostly focus on the scalability, low power communication, and applications. For example,  \cite{yousuf2018throughput}, and \cite{haxhibeqiri2017LoRa} focus on the scalability of LoRa. \cite{liando2019known}, \cite{xu2021pyramid}, \cite{chen2021lofi} and\cite{hu2020scLoRa} test the LoRa performance in throughout capacity, concurrency and coverage. PLoRa \cite{peng2018pLoRa} aims at utilize ambient backscatter for enabling LoRa communication. Other LoRa research works are mostly system applications in sensing field \cite{zhang2020exploring}.
\par The problem of CTI \cite{gollakota2011clearing} has been examined in the unlicensed spectrum, while they focus on communication protocols such as Wi-Fi, ZigBee and Bluetooth. Among them, Forward Error Correction(FEC) \cite{hithnawi2016crosszig} \cite{marcelis2020dare} add redundancy information for combating the errors due to interference, while they commonly suffer from limited error correction capability. On the other hand, interference Cancellation \cite{yubo2013zimo} \cite{zheng2014zisense} access the PHY level samples for demodulating the corrupted signal. Although these techniques are effective, they inevitably require a high sampling rate and high complexity, incurring significant costs on the low-cost and low-power LoRa devices.
\par In contrast, this paper introduces the first partial symbol recovery design for enhancing CTI resilience on LPWAN techniques, e.g., LoRa. The LoRa system achieves higher robustness by expanding the SF and coding rate. The adjustment of the two parameters of SF and CR fails to work when colliding with high power and dense CTI. On the one hand, the SF and CR in LoRa cannot be expanded infinitely, and the maximum SF of the existing COT LoRa device is set to 12, and the CR is 4/8. On the other hand, due to the difference in the duration of LoRa and CTI, one LoRa symbol could be interfered by multiple CTI packets, causing most of the demodulation errors of the PHY layer symbol. To confirm our statement, we measured the usage of the 2.4G Hz ISM band. A Wi-Fi AP sends 46940 packets every minute, and 95.7\% of the packets are less than 0.2ms. The shortest LoRa symbol has a duration of 0.625 ms in SF7. Different from existing works, our PHY layer elaborately examines the unique features of LoRa, so that it manages to recover corrupted LoRa symbols without a higher sampling rate. 

\section{Conclusion}
With emerging LPWAN techniques in the unlicensed spectrum, CTI inevitably corrupts this low-power communication. 
This paper presents the first partial symbol recovery design for improving the resilience of LPWAN against the high-power CTI. 
Different from existing works focusing on traditional wireless technologies, our PHY design relies on the unique features of LoRa communication for effective symbol recovery. Under the challenge of low SNR and multiple CTI, our PSR has a good performance. Extensive evaluation on USRP under the mainstream CTI sources, e.g., Wi-Fi, ZigBee and Bluetooth, demonstrate the reliability of our PSR across various settings.

\section{Acknowledgement}
This work was supported in part by National Natural Science Foundation of China under Grant No. 61902066, Natural Science Foundation of Jiangsu Province under Grant No. BK20190336, China National Key R\&D Program 2018YFB2100302 and Fundamental Research Funds for the Central Universities under Grant No. 2242021R41068.

\bibliographystyle{IEEEtran}

\bibliography{cross_LoRa_reference}

\end{document}